\RequirePackage{fix-cm}
\documentclass[twocolumn,epjc3,showpacs,preprintnumbers,amsmath,amssymb]{svjour3}
\smartqed  
\RequirePackage{graphicx}
\RequirePackage[numbers,sort&compress]{natbib}
\usepackage{graphicx,amsmath,amsfonts,latexsym,amssymb,graphics,epsfig,subfigure,color,makeidx,hyperref}
\usepackage{multirow}

\journalname{Eur. Phys. J. C}
\newcommand {\nn}    {\nonumber}
\begin{document}

\title{Complexity growth rates for AdS black holes in massive gravity and $f(R)$ gravity}

\author{Wen-Di Guo\thanksref{e1,addr1}
                \and Shao-Wen Wei\thanksref{e2,addr1}
             \and Yan-Yan Li\thanksref{e3,addr1}
             \and  Yu-Xiao Liu\thanksref{e4,addr1}
        }
\thankstext{e1}{e-mail:guowd14@lzu.edu.cn}
\thankstext{e2}{e-mail:weishw@lzu.edu.cn}
\thankstext{e3}{e-mail:liyy2015@lzu.edu.cn}
\thankstext{e4}{e-mail:liuyx@lzu.edu.cn, Corresponding author}

   \institute{
   Institute of Theoretical Physics, Lanzhou University, Lanzhou 730000, People's Republic of China \label{addr1}}
\date{Received: date / Accepted: date}

 \maketitle

\begin{abstract}

The ``complexity = action" duality states that the quantum complexity is equal to the action of the stationary AdS black holes within the Wheeler-DeWitt patch at late time approximation. We compute the action growth rates of the neutral and charged black holes in massive gravity and the neutral, charged and Kerr-Newman black holes in $f(R)$ gravity to test this conjecture. Besides, we investigate the effects of the massive graviton terms, higher derivative terms and the topology of the black hole horizon on the complexity growth rate.

\end{abstract}

\maketitle

\section{Introduction}

The quantum computational complexity was recently proposed by Susskind and his collaborators \cite{Susskind:2013aaa,Susskind:2014rva,Susskind:2016tae,Brown:2015bva,Brown:2015lvg} for the first time. They connected the black hole interior  with the quantum computational complexity, which was defined as the minimal number of elementary operations (also named quantum gates) on constructing the boundary quantum state from a reference quantum state \cite{Brown:2015bva,Brown:2015lvg}. This is a remarkable progress on understanding black hole interior. After that, the quantum complexity has been extensively investigated \cite{Hosur:2015ylk,Alishahiha:2015rta,Lehner:2016vdi,Barbon:2015ria,Carmi:2016wjl,Chapman:2016hwi,
Ben-Ami:2016qex,Brown:2016wib,Roberts:2016hpo,Couch:2016exn,Barbon:2015soa,Reynolds:2016rvl,
Yang:2016awy,Brown:2017jil,Huang:2016fks,Zhao:2017iul,Tao:2017fsy,Yangfr:2017,Cai:2017sjv,Momeni:2016ira,
Mazhari:2016yng,Momeni:2016qfv,Momeni:2016ekm,Momeni:2017ibg}. The gravitational action with null boundaries was proposed and discussed in detail in Ref.~\cite{Lehner:2016vdi}, and the  connection between the second law of thermodynamics with the quantum complexity was investigated in Ref.~\cite{Brown:2017jil}.

Stanford and Susskind first related the complexity $\mathcal C$ with the spatial volume $V$ of the Einstein-Rosen bridge which connects two boundaries as \cite{Stanford:2014jda}:
\begin{eqnarray}
\mathcal C\sim\frac{V}{GL},
\end{eqnarray}
where $G$ is the Newton constant and $L$ is a length scale that should be chosen to be either the anti de-Sitter (AdS) radius or the radius of the black hole horizon. This conjecture named as Complexity--Volume (CV) duality has a lot of nice features and it has been tested in the  shock wave geometries \cite{Brown:2015lvg}. It is worthwhile noting that the length scale $L$ should be chosen  according to the explicit situation. The subsequent Complexity--Action (CA) duality does not involve any ambiguous quantities and preserves all the good features of the CV duality \cite{Brown:2015lvg}. The CA duality relates the quantum complexity to the action of the black hole in the  Wheeler-DeWitt (WDW) patch:
\begin{eqnarray}
\mathcal C=\frac{A}{\pi \hbar}.
\end{eqnarray}
The authors of Ref.~\cite{Couch:2016exn} revisited the connection between the thermodynamical volume of the black hole and the action growth rate in the WDW patch at late time approximation.
The UV divergence of the action in that region was investigated in Refs.~\cite{Reynolds:2016rvl,Carmi:2016wjl}.
It was pointed out by Lloyd that the quantum complexity growth rate is bounded by \cite{Lloyd2000}
\begin{eqnarray}
\frac{d \mathcal C}{dt}\leq\frac{2E}{\pi\hbar},
\end{eqnarray}
where $E$ is the average energy of the quantum state relating to the ground state. If the CA duality is correct, the action growth rate of the black holes in the WDW patch will also be bounded. This was tested by computing the action growth rate. And some concrete forms of the action growth rate of the AdS black holes in the WDW patch at the late time approximation were proposed \cite{Brown:2015bva,Brown:2015lvg}.

However, the authors of Ref.~\cite{Cai:2016xho} found that the above bound will be violated for both small and large  charged static AdS black holes, so they proposed a new bound
\begin{eqnarray}\label{conjectureofcai}
\frac{dA}{dt}\leq(M-\Omega_+J-\mu_+Q)-(M-\Omega_-J-\mu_-Q),
\end{eqnarray}
where $\Omega$ and $\mu$ denote respectively the angular velocity and chemical potential of the black holes, and the parameters $M$, $J$, and $Q$ are the black hole mass, angular momentum, and charge, respectively. The subscript $+$ ($-$) denotes the quantities evaluated at the outer (inner) horizon. They conjectured that all stationary AdS black holes could saturate this bound for gravity theories without higher derivative terms of curvature. As a test, the action growth rates of $D$-dimensional Reissner Nordstr\"{o}m-AdS black holes, Kerr-AdS black holes, rotating and charged BTZ black holes in general relativity (GR), as well as the charged Gauss-Bonnet-AdS black holes have been calculated. This conjecture was generalized to be a compact form in Ref.~\cite{Huang:2016fks}. It was shown that the difference of the generalized enthalpy between the two corresponding horizons decides the action growth rate~\cite{Huang:2016fks} and strong energy conditions of the matter outside the horizon of the black holes can ensure the validity of the complexity growth rate bound \cite{Yang:2016awy}. The action growth rates of the charged black holes with a single horizon were also studied in Ref. \cite{Cai:2017sjv}. The Lloyd bounds of the action growth rates of black holes in Born-Infeld gravity and in $f(R)$ gravity were also tested in Refs. \cite{Tao:2017fsy,Yangfr:2017} by considering the contributions from the null surface and the joints. They found that the Lloyd bounds for charged black holes will usually be violated in both Born-Infeld gravity and $f(R)$ gravity. \textcolor[rgb]{0.00,0.50,1.00}{Besides, the action growth rates based on the CA conjecture were also studied in massive gravity and higher derivative theories in Refs. \cite{Pan:2016ecg,Alishahiha:2017hwg}.} In this paper, we only focus on the conjecture \eqref{conjectureofcai} in this paper rather than the Lloyd bound because the conjecture \eqref{conjectureofcai} is based on the violation of the Lloyd bound.

This paper is organized as follows. In Sec. \ref{actiongrowthrateinGR}, we first verify the KN case in GR. In Sec. \ref{actionofmassivegravity} we calculate the action growth rate of the neutral and charged black holes in higher dimensional massive gravity within the WDW patch at the late time approximation, and study the influence of the massive graviton terms. Besides, whether the parameter $k$ which denotes the topology of the black hole horizon will affect the action growth rate is studied. In Sec. \ref{actionoffR} the action growth rate of the black holes in $f(R)$ gravity is studied, for the cases of neutral, charged, and KN black holes. Finally, the conclusions are given in Sec.\ref{conclusion}.

\section{Action growth rate of Kerr-Newman black holes in GR }\label{actiongrowthrateinGR}

Before dealing with the massive gravity and $f(R)$ gravity, we start with the KN black holes in GR. The action is given by
\begin{eqnarray}
\mathcal A&=&\mathcal A_{bk}+\mathcal A_{bd}\nn\\
          &=&\frac{1}{16 \pi G}\int_M d^4x\sqrt{-g}\left(R-2 \Lambda-G F_{\mu \nu } F^{\mu \nu } \right)\nn\\
          &&+\frac{1}{8 \pi G}\int _{\partial M}d^3x\sqrt{-h}K,
\end{eqnarray}
where $\mathcal A_{bk}$ and $\mathcal A_{bd}$ denote the actions of the bulk and boundary terms, respectively, and $h$ is the induced metric of the hypersurface and $K$ is the trace of the extrinsic curvature. The solutions of the metric and the electric potential are given in Refs. \cite{Carter:1968ks,Plebanski:1976gy}:
\begin{eqnarray}\label{KNmetric}
ds^2&=&-\left(\frac{\triangle_r}{\rho^2}-\frac{\triangle_\theta \sin^2\theta}{\rho^2}a^2\right)dt^2
     +\frac{\rho^2}{\triangle_r}dr^2
     +\frac{\rho^2}{\triangle_\theta}d\theta^2\nn\\
     &&+2\frac{a\triangle_r \sin^2\theta-a(r^2+a^2)\triangle_\theta \sin^2 \theta}{\rho^2\Xi}dtd\phi\\
     &~&+\frac{\sin^2\theta}{\rho^2\Xi^2}
     \left[(r^2+a^2)^2\triangle_\theta-a^2\triangle_r \sin^2\theta \right]d\phi^2,\nn\\
A&=& -\frac{qr}{\rho ^2}dt+\frac{ qra \sin^2\theta}{\Xi  \rho ^2}d\phi.
\end{eqnarray}
Here we adopt the conventions of Ref. \cite{Caldarelli:1999xj}. The metric functions are given by
\begin{eqnarray}
&&\triangle _r=\left(a^2+r^2\right) \left(1+\frac{r^2}{l^2}\right)-2 Gmr+Gq^2,\nn\\
&&\triangle _{\theta }=1-\frac{a^2 \cos ^2\theta}{l^2},
~~~~~\rho ^2=a^2 \cos ^2\theta+r^2,\nn\\
&&\Xi =1-\frac{a^2}{l^2},
\end{eqnarray}
where $a$ is the rotational parameter. Note that, this solution is valid only for $a^2<l^2$. The thermodynamical quantities are \cite{Caldarelli:1999xj}
\begin{eqnarray}
&&M=\frac{m}{\Xi ^2},~J=\frac{a m}{\Xi ^2},~Q=\frac{q}{\Xi },\nn\\
&&\mu _{\pm }=\frac{q r_{\pm }}{\left(r_{\pm }{}^2+a^2\right)},~\Omega _{\pm }=\frac{a \Xi }{r_{\pm }{}^2+a^2}.\label{ther}
\end{eqnarray}
Solving $\triangle_r(r_{\pm})=0$, we can get
 \begin{eqnarray}
 m&=&\frac{\left(r_-+r_+\right) \left(a^2+l^2+r_-^2+r_+^2\right)}{2 G l^2},\nn\\
 q^2&=&\frac{r_- r_+ \left(l^2+r_-^2+r_+^2+r_- r_+\right)}{G l^2}\nn\\
    &+&a^2 \frac{\left(r_- r_+-l^2\right)}{G l^2}.
 \end{eqnarray}
 The determinant of this metric is
\begin{eqnarray}
\sqrt{-g}=\frac{\sin \theta}{\Xi }\rho^2.
\end{eqnarray}

The Ricci scalar $R$ and the field strength $F^2$ can be calculated straightforwardly as
\begin{eqnarray}
R&=&-\frac{12}{l^2},\nn\\
F^2&=&-4 q^2\Big(\frac{ a^4 \cos (4 \theta )+3 a^4+4 a^2 \left(a^2-6 r^2\right) \cos (2 \theta )}{\left(a^2 \cos (2 \theta )+a^2+2 r^2\right)^4}\nn\\
&&+\frac{8 r^4-24 a^2 r^2}{\left(a^2 \cos (2 \theta )+a^2+2 r^2\right)^4}\Big).
\end{eqnarray}
Then we can get the action growth rate of the Einstein-Maxwell part
\begin{eqnarray}
\frac{d\mathcal A_{bk}}{dt}&=&\frac{2\pi}{16 \pi G}\int_{r_-}^{r_+}\int_0^\pi
                        \sqrt{-g}\left(R-2\Lambda-G F_{\mu\nu}F^{\mu\nu}\right)d\theta dr\nn\\
                      &=&\frac{1}{2} (r_+-r_-) \Bigg(\frac{l^2 q^2 \left(a^2-r_- r_+\right)}{\left(a^2- l^2\right) \left(a^2+r_-^2\right) \left(a^2+r_+^2\right)}\nn\\
                      &&+\frac{a^2+r_-^2+r_-r_++r_+^2}{G (a^2 - l^2)}\Bigg).
\end{eqnarray}
\textcolor[rgb]{0.00,0.50,1.00}{Note that, we integral the action from the inner horizon $r_-$ to the outer horizon $r_+$ because other regions of the WDW patch do not contribute to the action growth rate at late time approximation. The contribution of the region outside the horizon is independent on time and that of the region behind the past horizon shrinks exponentially to zero \cite{Brown:2015lvg}.}
The trace of the extrinsic curvature $K$ can be gotten as
\begin{eqnarray}
K&=&\frac{1}{\sqrt{-g}}\partial _r\left(\frac{\sqrt{-g}}{\rho} \sqrt{\triangle_r}\right)\nn\\
 &=&\frac{ r \sqrt{\triangle_r}}{ \rho ^3 }+\frac{\partial_r \triangle_r}{2\rho \sqrt{\triangle_r}}.
\end{eqnarray}
So the contribution from the boundary term will be
\begin{eqnarray}
\frac{d\mathcal A_{bd}}{dt}&=&\frac{\partial_r \triangle r}{4 G \Xi }
                               \bigg|^{r_+}_{r_-}\nn\\
                            &=&\left(r_+-r_-\right)\frac{ a^2+l^2+2 \left(r_-^2+r_+ r_-+r_+^2\right)}{2 G \left(l^2-a^2\right)},
\end{eqnarray}
where we have used $\triangle_r (r_{\pm})=0$, and the angle part has been integrated. In the end, we can get the action growth rate of a KN AdS black hole in GR
\begin{eqnarray}
\frac{d\mathcal A}{dt}&=&\left(r_+-r_-\right)\Bigg(\frac{ a^4 \left(2 l^2+r_-^2+r_+^2\right)}{2 G \left(l^2-a^2\right) \left(a^2+r_-^2\right) \left(a^2+r_+^2\right)}\nn\\
&&+\frac{a^2 \left(l^2 \left(r_--r_+\right){}^2+\left(r_-^2+r_+^2\right){}^2\right)}{2 G \left(l^2-a^2\right) \left(a^2+r_-^2\right) \left(a^2+r_+^2\right)}\nn\\
&&+\frac{2 r_-^2 r_+^2 \left(l^2+r_-^2+r_+^2+r_- r_+\right)}{2 G \left(l^2-a^2\right) \left(a^2+r_-^2\right) \left(a^2+r_+^2\right)}\Bigg).
\end{eqnarray}
With the help of the thermodynamic quantities (\ref{ther}), we can rewrite it in the following form
\begin{eqnarray}
\frac{d\mathcal A}{dt}=\left(M-\Omega _+J - \mu _+Q\right)-\left(M- \Omega _-J- \mu _-Q\right).
\end{eqnarray}
Clearly, this is just the action growth rate of a KN AdS black hole in GR speculated in Ref. \cite{Cai:2016xho}.

\section{Action growth rate of black holes in massive gravity}
\label{actionofmassivegravity}

We first give a brief review of the black hole solutions in  $(n+2)$-dimensional massive gravity \cite{deRham:2010ik,deRham:2010kj,Hinterbichler:2011tt,Hassan:2011hr,Hassan:2011tf}. The action is given by \cite{Vegh:2013sk,Cai:2014znn}
\begin{eqnarray}
\mathcal A&=&\mathcal A_{bk}+\mathcal A_{bd}\nn\\
          &=&\frac{1}{16 \pi G}\int_M d^{n+2}x\sqrt{-g}\Bigg(R+\frac{(n+1) n}{l^2}-G F^2+\nn\\
          &&m^2 {\sum _{i=1}^4 c_i \mathcal{U}_i(g,\hat f)}\Bigg)+\frac{1}{8\pi G}\int_{\partial M} d^{n+1}x\sqrt{-h}K, \label{Action}
\end{eqnarray}
where $g$, $\hat f$, and $m$ are respectively the spacetime metric, the reference metric, and the mass of graviton. The coefficients $c_i$ are negative constants if one requires $m^2>0$ \cite{Cai:2014znn}, and $\mathcal{U}_i$ are defined by the following symmetric polynomials:
\begin{eqnarray}
\mathcal{U}_1&=&[\mathcal{K}], \nn\\
\mathcal{U}_2&=&[\mathcal{K}]^2-[\mathcal{K}^2], \nn\\
\mathcal{U}_3&=&[\mathcal{K}]^3-3[\mathcal{K}][\mathcal{K}^2]+2[\mathcal{K}^3], \nn\\
\mathcal{U}_4&=&[\mathcal{K}]^4-6[\mathcal{K}^2][\mathcal{K}]^2+8[\mathcal{K}^3][\mathcal{K}]
                +3[\mathcal{K}^2]^2-6[\mathcal{K}^4],
\end{eqnarray}
where $[X]$ denotes the trace of the matrix $X$. The matrix $\mathcal{K}^\mu_{~~\nu}$ is defined as $\mathcal{K}^\mu_{~~\nu}=\sqrt{g^{\mu\alpha}\hat f_{\alpha\nu}}$. Varying the action with respect to the metric $g_{\mu\nu}$ and the gauge field $A_\mu$ we can get the
field equations:
\begin{eqnarray}\label{fieldequation}
&&R_{\mu\nu}-\frac{1}{2}Rg_{\mu\nu}-\frac{1}{2l^2}{n(n+1)}g_{\mu\nu}+m^2\chi_{\mu\nu}\nn\\
&&=2G \left(F_{\mu\sigma}F_\nu^{~~\sigma}-\frac{1}{4} g_{\mu\nu}F^2\right), \nn\\
&&\nabla_\mu F^{\mu\nu}=0,
\end{eqnarray}
where the mass term $\chi_{\mu\nu}$ is given by
\begin{eqnarray}
\chi_{\mu\nu}=&-&\frac{c_1}{2}\big(\mathcal{U}_1g_{\mu\nu}-\mathcal{K}_{\mu\nu}\big)\nn\\
              &-&\frac{c_2}{2}\big(\mathcal{U}_2g_{\mu\nu}-2\mathcal{U}_1\mathcal{K}_{\mu\nu}+2\mathcal{K}^2_{\mu\nu}\big)\nn\\
              &-&\frac{c_3}{2}\big(\mathcal{U}_3g_{\mu\nu}-3\mathcal U_2\mathcal K_{\mu\nu}
              +6\mathcal U_1\mathcal K^2_{\mu\nu}-6\mathcal K^3_{\mu\nu}\big)\nn\\
              &-&\frac{c_4}{2}\big(\mathcal U_4g_{\mu\nu}-4\mathcal U_3\mathcal K_{\mu\nu}+12\mathcal U_2\mathcal K^2_{\mu\nu}\nn\\
              &-&24\mathcal U_1\mathcal K^3_{\mu\nu}+24\mathcal K^4_{\mu\nu}\big).
\end{eqnarray}
The solution for a static black hole is \cite{Cai:2014znn}:
\begin{eqnarray}\label{metric}
 ds^2&=&-f(r)dt^2+f^{-1}(r)dr^2+r^2\hat{h}_{ij}dx^i dx^j,\nn\\
 && (i,j=1,2,3,\cdots,n)\\ \label{referencemetric}
 \hat f_{\mu\nu}&=&\text{diag}(0,0,\hat{h}_{ij}),\\
 A_\mu &=& (A_t,0,0,\cdots,0),
\end{eqnarray}
where $\hat{h}_{ij}$ is the induced metric of the hypersurface space. The explicit forms of $f(r)$ and $A_t$ are \cite{Cai:2014znn}
\begin{eqnarray}
f(r)&=&k+\frac{r^2}{l^2}-\frac{2G m_0}{r^{n-1}}+\frac{2G q^2}{n(n-1)r^{2(n-1)}}\nn\\
     &&+\frac{c_1m^2}{n}r+c_2m^2+\frac{(n-1)c_3m^2}{r} \nn\\
     &&+\frac{(n-1)(n-2)c_4m^2}{r^2},\\
  A_t &=&     \frac{q}{(n-1)r^{n-1}} - \frac{q}{(n-1)r_+^{n-1}},
\end{eqnarray}
where $k=-1,0,1$ denotes the topology of the black hole horizon, corresponding to hyperbolic, Ricci flat, or spherical, respectively. The physical mass $M$ and charge $Q$ of the black hole are given by
\begin{eqnarray}
M&=&\frac{nV_n}{8\pi }m_0,\label{mass} \\
Q&=&\frac{V_n}{4\pi }q,\label{charge}
\end{eqnarray}
where $V_n$ is the volume of the $n$-dimensional space that depends on the topology of the black hole horizon.
The $P$-$V$ criticality and the	Van der Waals like behavior of the massive gravity black holes have been investigated in Refs. \cite{Xu:2015rfa,Hendi:2017fxp,Hendi:2016yof,Hendi:2015hoa,Hendi:2015eca,Cai:2012db,Zou:2016sab,Zou:2017juz}.

Next we will investigate the effects of the mass of graviton, the dimension of spacetime, and the topology of the black hole horizon on the action growth rates for both the neutral and charged massive black holes.

\subsection{Action growth rate of the neutral black holes in higher dimensional massive gravity}

For neutral black holes, there is no Maxwell field part in the action.
With the metric \eqref{metric}, we can calculate the Ricci scalar $R$ for the neutral black holes:
\begin{eqnarray}
R=&-&\frac{(n+1) (n+2)}{l^2}-\frac{c_1 m^2 (n+1)}{r}-\frac{c_2 m^2 n
      (n-1)}{r^2} \nn\\
  &-&\frac{c_3 m^2 (n-1)^2 (n-2)}{r^3}\nn\\
  &-&\frac{c_4 m^2 (n-1) (n-2)^2 (n-3)}{r^4}.
\end{eqnarray}
Following Refs.~\cite{Brown:2015lvg,Cai:2016xho}, the action growth rate of the bulk action in the WDW patch at late time approximation limit is
\begin{eqnarray}
\frac{d \mathcal A_{bk}}{dt}&=&\frac{V_n}{16\pi G} \int_{0 }^{r_h} r^n \Big(R+\frac{(n+1) n}{l^2}\nn\\
                &&+m^2 {\sum _{i=1}^4 c_i \mathcal{U}_i(g,f)}\Big) \, dr\nn\\
              &=&\frac{V_n}{16\pi G}\Bigg(-m^2\frac{c_1}{n}r_h^n+m^2c_3(n-1)r_h^{n-2}\nn\\
              &&+2m^2c_4(n-1)(n-2)r_h^{n-3}-\frac{2}{l^2}r^{n+1}
                                       \Bigg),
\end{eqnarray}
where $r_h$ denotes the horizon radius.
Next, let us turn to the YGH boundary term. The trace of the extrinsic curvature with the metric \eqref{metric} is
\begin{eqnarray}
K&=&\frac{1}{r^n}\partial_r(r^n\sqrt f)\nn\\
 &=&\frac{n}{r}\sqrt f+\frac{f'}{2\sqrt f},
\end{eqnarray}
where the prime denotes the derivative with respect to $r$. So the action growth rate of the surface term in the WDW patch is
\begin{eqnarray}
\frac{d\mathcal A_{bd}}{dt}&=&\frac{V_n}{8\pi G}\bigg[r^n\sqrt f(\frac{n}{r}\sqrt f+\frac{f'}
                       {2\sqrt f})\bigg]^{r_h}_0\nn\\
                   &=&\frac{V_n}{8\pi G}\bigg(nr_h^{n-1}k+\frac{n+1}{l^2}r_h^{n+1}\nn\\
                     &&  +(1+\frac{1}{2n})c_1m^2r_h^n+nc_2m^2r_h^{n-1}\nn\\
                   &~&+(n-\frac{1}{2})(n-1)c_3m^2r_h^{n-2}\nn\\
                      && +(n-1)^2(n-2)c_4m^2r_h^{n-3}
                   \bigg).
\end{eqnarray}
Then we can get the total action growth rate of the higher dimensional massive neutral black holes in the WDW patch at the late time approximation limit:
\begin{eqnarray}
\frac{d\mathcal A}{dt}&=&\frac{d \mathcal A_{bk}}{dt}+\frac{d\mathcal A_{bd}}{dt}\nn\\
             &=&\frac{n V_n}{8\pi G}\bigg(k r_h^{n-1}+\frac{r_h^{n+1}}{l^2}+\frac{1}{n}c_1m^2r_h^n\nn\\
                                        &~&+c_2m^2r_h^{n-1}+(n-1)c_3 m^2r_h^{n-2}\nn\\
                                        &&+(n-1)(n-2)c_4m^2r_h^{n-3}
                                    \bigg).
\end{eqnarray}
Solving $f(r_h)=0$, we get
\begin{eqnarray}\label{m0}
m_0&=&\frac{1}{2G}\Big(kr_h^{n-1}+\frac{r_h^{n+1}}{l^2}+\frac{1}{n}c_1m^2r_h^{n}\nn\\
    &~&+c_2m^2r_h^{n-1}+(n-1)c_3m^2r_h^{n-2}\nn\\
    &&+(n-1)(n-2)c_4m^2r_h^{n-3}\Big).
\end{eqnarray}
Combining with Eq. \eqref{mass}, we can get the same result with Refs.~\cite{Pan:2016ecg,Cai:2016xho}:
\begin{eqnarray}
\frac{d\mathcal A}{dt}=2M.
\end{eqnarray}
If the CA conjecture is correct, then we will get
\begin{eqnarray}
\frac{d\mathcal C}{dt}=\frac{2M}{\pi\hbar}.
\end{eqnarray}
The result is independent of the spacetime dimension, the mass of graviton, and the topology of the black hole. If a neutral stationary AdS black hole in massive gravity has the same mass as the black hole in GR, they will have the same growth rate of complexity. So we conclude that the neutral massive black holes will also be the fastest computer.

\subsection{Action growth rate of the charged black holes in higher dimensional massive gravity}

For the case of a charged black hole, the field strength of the electromagnetic field turns out to be
\begin{eqnarray}
F^2=-2 q^2 r^{-2 n},
\end{eqnarray}
and the parameters $m_0$ and $q$ can be solved from $f(r_\pm)=0$,
\begin{eqnarray}
m_0&=&\frac{r_+r_-^{2n-1}}{2G(r_+r_-^n-r_-r_+^n)}
      \bigg(k+\frac{r_-^2}{l^2}+\frac{c_1}{n}m^2r_{-}+c_2m^2\nn\\
            &&+(n-1)c_3m^2\frac{1}{r_-}+(n-1)(n-2)c_4 m^2\frac{1}{r_-^2}\bigg)\nn\\
  &&-\frac{r_+^{2n-1}r_-}{2G(r_+r_-^n-r_-r_+^n)}
      \bigg(k+\frac{r_+^2}{l^2}+\frac{c_1}{n}m^2r_{+}+c_2m^2\nn\\
           && +(n-1)c_3m^2\frac{1}{r_+}+(n-1)(n-2)c_4 m^2\frac{1}{r_+^2}\bigg),\nn\\
q^2&=&\frac{(n-1)}{2G}\bigg(n k {r}_-^{n-1} {r}_+^{n-1}+\frac{n
\left({r}_-^{2n}{r}_+^{n-1}-{r}_-^{n-1}{r}_+^{2n}\right)}{l^2 \left( {r}_-^{n-1}- {r}_+^{n-1}\right)}\nn\\
&&-\frac{c_1m^2(r_-^{n}-r_+^n)}{r_-^{1-n}-r_+^{1-n}}
+n c_2 m^2 {r}_+^{n-1}{r}_-^{n-1}\nn\\
&&+\frac{nc_3m^2(r_-^{n-2}-r_+^{n-2})}{r_+^{1-n}-r_-^{1-n}}\nn\\
&&+
\frac{(n-1)(n-2)c_4m^2(r_-^{n-3}-r_+^{n-3})}{r_+^{1-n}-r_-^{1-n}}\bigg).
\end{eqnarray}
The bulk action growth rate of the charged massive black holes within the WDW patch at the late time approximation limit can be calculated straightforwardly:
\begin{eqnarray}
\frac{d\mathcal A_{bk}}{dt}&=&\frac{ V_n}{8\pi G}\bigg[k\left({r}_+^{n-1}-{r}_-^{n-1}\right)+\frac{c_1}{2 n}m^2  \left({r}_+^n-{r}_-^n\right)\nn\\
&&+c_2 m^2  \left({r}_+^{n-1}-{r}_-^{n-1}\right)
\nn\\&&+\frac{3 (n-1)c_3}{2 }  m^2 \left({r}_+^{n-2}- {r}_-^{n-2}\right)\nn\\
&~&
+2 c_4 m^2 (n-2) (n-1)  \left({r}_+^{n-3}-{r}_-^{n-3}\right)\bigg].
\end{eqnarray}
Besides, the contribution from the YGH surface term within the WDW patch is
\begin{eqnarray}
\frac{d\mathcal A_{bd}}{dt}&=&
   \frac{V_n}{16\pi G}\Big[2k (n-1) \left({r}_+^{n-1}-{r}_-^{n-1}\right)\nn\\
   &&+ c_1 m^2 \frac{(2 n-1)}{n} \left({r}_+^n-{r}_-^n\right)\nn\\
   &&+ 2c_2 m^2  (n-1) \left({r}_+^{n-1}-{r}_-^{n-1}\right) \nn\\
&&+c_3 m^2  (n-1) (2 n-3) \left({r}_+^{n-2}-{r}_-^{n-2}\right)\nn\\
  &&+2c_4 m^2  (n-2)^2 (n-1) \left({r}_+^{n-3}-{r}_-^{n-3}\right)\Big].
\end{eqnarray}
Therefore, the total action growth rate of the charged massive black holes in the WDW patch at the late time approximation limit is
\begin{eqnarray}\label{agocmbh}
\frac{d\mathcal A}{dt}&=&\frac{d\mathcal A_{bk}}{dt}+\frac{d\mathcal A_{bd}}{dt}\nn\\
             &=&\frac{4\pi}{(n-1)V_n}Q^2\left(\frac{1}{r_-^{n-1}}-\frac{1}{r_+^{n-1}}\right) \nn\\
             &=&\mu _- Q-\mu _+ Q,\label{aa}
\end{eqnarray}
where we have used the charge of the black hole \eqref{charge}, and the chemical potential is defined as follows
\begin{eqnarray}
\mu_{\pm} =\frac{q}{(n-1) r_{\pm}^{n-1}}.
\end{eqnarray}
From Eq. \eqref{aa}, we can see that, no matter what the value $k$ takes, the result is consistent with that of Ref. \cite{Cai:2016xho}. In other words, the topology of the charged black hole horizon will not affect the quantum complexity growth rate. However, as the authors of \cite{Pan:2016ecg} showed, when the mass $M$ and charge $Q$ of the massive gravity black holes are the same as that of the black holes in GR, the complexity of the charged black holes in massive gravity theory will grow faster than that of the charged black holes in GR. The reason is as follows. If we define the quantities in GR with a hat, then we will get
\begin{eqnarray}
\hat f(r)-f(r)&=& - m^2\Big(\frac{c_1}{n}r+c_2+\frac{(n-1)c_3}{r}\nn\\&&
                  +\frac{(n-1)(n-2)c_4}{r^2}\Big)>0,
\end{eqnarray}
for negative parameters $c_i$, which will ensure the positivity of the mass square of graviton. So we have the following relations
\begin{eqnarray}
 r_-<\hat r_- ~~~~ \text{and}~~~~ r_+>\hat r_+.
\end{eqnarray}
Having
\begin{eqnarray}
\frac{d\mathcal C}{dt}=\frac{2\kappa^2}{V_n}Q^2\left(\frac{1}{r_-^{n-1}}-\frac{1}{r_+^{n-1}} \right),
\end{eqnarray}
and
\begin{eqnarray}
\frac{d\hat{\mathcal C}}{dt}=\frac{2\kappa^2}{V_n}Q^2\left(\frac{1}{\hat r_-^{n-1}}-\frac{1}{\hat r_+^{n-1}} \right),
\end{eqnarray}
one has
\begin{eqnarray}
\frac{d\mathcal C}{dt}>\frac{d\hat{\mathcal C}}{dt}.
\end{eqnarray}
So, the charged black holes in massive gravity are faster computers than the charged black holes in GR, though the bound \eqref{conjectureofcai} is not exceeded.

\section{Action growth rate of black holes in $f(R)$ gravity}\label{actionoffR}
The action for the $f(R)-$Maxwell gravity is
\begin{eqnarray}\label{action of f(R)}
\mathcal A&=&\mathcal A_{bk}+\mathcal A_{bd}\nn\\
          &=&\frac{1}{16\pi G}\int_M d^4x\sqrt{-g}\left[R +f (R)-GF^2\right]\nn\\
          &&+\frac{1}{8\pi G}\int _{\partial M}d^3x\sqrt{-h}\left[ \left(1+ f'(R) \right)K\right],
\end{eqnarray}
where the prime denotes the derivative with respect to $R$, and the YGH surface term is different from that of in GR \cite{Guarnizo:2010xr}.
The static black hole solution was found in Ref. \cite{Moon:2011hq}. The authors assumed the following metric form
\begin{eqnarray}\label{metricfR}
ds^2=-N(r)dt^2+\frac{dr^2}{N(r)}+r^2d\Omega_2^2,
\end{eqnarray}
with the metric function given by
\begin{eqnarray}
N(r)=1-\frac{2Gm}{r}+\frac{Gq^2}{(1+f'(R_0))r^2}-\frac{R_0}{12}r^2.
\end{eqnarray}
Here the parameters $m$ and $q$ are related to the physical mass and charge
\begin{eqnarray}\label{physicalquantites}
M&=&(1+f'(R_0))m,\nn\\
Q&=&\frac{q}{\sqrt{1+f'(R_0)}}.
\end{eqnarray}
According to Ref. \cite{Sheykhi:2012zz} $R_0$ is the constant scalar curvature
\begin{eqnarray}\label{constant R}
R_0=\frac{2f(R_0)}{f'(R_0)-1}=4\Lambda=-\frac{12}{l^2},
\end{eqnarray}
where $\Lambda$ is the cosmology constant and $l$ is the AdS radius. Note that, in order to get constant curvature, the trace of the energy-momentum tensor of the Maxwell field should be zero, in other words, the spacetime should be four dimensional \cite{Sheykhi:2012zz}. The other solutions and thermodynamical qualities for $f(R)$ gravity black holes have also been investigated in Refs. \cite{Chen:2013ce,delaCruzDombriz:2012xy,Cembranos:2011sr,Wang:2013nvz,Chakraborty:2014xla,
Chakraborty:2016ydo,Hendi:2014mba,Hendi:2016oxk,Cai:2014upa,Cai:2013lqa}.
The authors of \cite{Cai:2016xho} conjectured that only the stationary black holes in gravity theories without higher derivative terms of curvature can saturate Eq.~\eqref{conjectureofcai}. So what the situation will be in $f(R)$ gravity?
\subsection{Action growth rate of the neutral black holes in $f(R)$ gravity}
In this section we will calculate the action growth rate of a static neutral black hole in $f(R)$ gravity within the WDW patch at the late time approximation limit.

We can get the mass parameter from $N(r_h)=0$:
\begin{eqnarray}\label{massoffR}
m=\frac{1}{2G}\left(r_h+\frac{r_h^3}{l^2}\right).
\end{eqnarray}
The contribution from the  bulk action  is
\begin{eqnarray}
\frac{d\mathcal A_{bk}}{dt}&=&\frac{4 \pi }{16 \pi  G}\int_{0}^{ r_h} r^2 (f(R_0)+R_0) \, dr\nn\\
                  &=&-\frac{1+f'(R_0)}{2G l^2}r_h^2.
\end{eqnarray}
The extrinsic curvature with the metric \eqref{metricfR} is
\begin{eqnarray}
K=\frac{2}{r}\sqrt N+\frac{N'}{2\sqrt N},
\end{eqnarray}
so the action growth rate of the YGH surface term in the WDW patch at the late time approximation limit is
\begin{eqnarray}
\frac{d\mathcal A_{bd}}{dt}&=&\frac{4\pi}{8\pi G}r^2(1+f'(R_0))\sqrt N K|^{r_h}_0\nn\\
                   &=&\frac{1+f'(R_0)}{2G}(2r_h+\frac{3}{l^2}r_h^3).
\end{eqnarray}
Thus the total action growth rate will be
\begin{eqnarray}
\frac{d\mathcal A}{dt}&=&\frac{d\mathcal A_{bk}}{dt}+\frac{d\mathcal A_{bd}}{dt}\nn\\
             &=&2(1+f'(R_0))m\nn\\
             &=&2M,
\end{eqnarray}
where we have used Eqs. \eqref{physicalquantites} and \eqref{massoffR}. The interesting thing is that,  in $f(R)$ gravity theory, the action growth rate equation \eqref{conjectureofcai} can also be saturated. That is to say, the higher derivative term of curvature in this theory does not change the action growth rate, so the neutral black holes in $f(R)$ gravity will still be the fastest computer. This result has been observed in Refs. \cite{Alishahiha:2017hwg,Yangfr:2017}.

\subsection{Action growth rate of charged black holes in $f(R)$ gravity}
The electromagnetic field strength is
\begin{eqnarray}
F^2=-\frac{2 q^2}{r^4}.
\end{eqnarray}
The parameters of the mass and charge in the function $N(r)$ can be solved from $N(r_{\pm})=0$:
\begin{eqnarray}
m&=&\frac{\left(r_-+r_+\right) \left(l^2+r_-^2+r_+^2\right)}{2 G l^2},\\
q^2&=&\frac{r_- r_+ \left(1 +f'(R_0)\right) \left(l^2+r_-^2+r_+^2+r_- r_+\right)}{G l^2}.
\end{eqnarray}
Then the bulk action growth rate of a charged black hole in $f(R)$ gravity within the WDW patch at the late time approximation limit can be calculated as
\begin{eqnarray}
\frac{d\mathcal A_{bk}}{dt}&=&\frac{\left(r_-^3-r_+^3\right) \left(1+f'(R_0)\right)}{2 G l^2}\nn\\
&&+\frac{1}{2} q^2 \left(\frac{1}{r_-}-\frac{1}{r_+}\right).
\end{eqnarray}
And the contribution from the YGH surface term is
\begin{eqnarray}
\frac{d\mathcal A_{bd}}{dt}&=&(1+f'(R_0)) \left(r_+-r_-\right)\nn\\
                           &&\times\frac{ 2 l^2+3 \left(r_-^2+r_+ r_-+r_+^2\right)}{2 G l^2}\nn\\
                           &&-\frac{q^2 \left(r_+-r_-\right)}{2 r_- r_+}.
\end{eqnarray}
Then we can get the total action growth rate,
\begin{eqnarray}
\frac{d\mathcal A}{dt}&=&\frac{d\mathcal A_{bk}}{dt}+\frac{d\mathcal A_{bd}}{dt}\nn\\
             &=&\frac{\left(r_+-r_-\right) \left(1+f'(R_0)\right) \left(l^2+r_-^2+r_+^2+r_-r_+\right)}{G l^2}\nn\\
             &=&\mu _- Q-\mu _+ Q,
\end{eqnarray}
where the chemical potentials at the inner and outer horizon are $\mu_{\pm}=\frac{q\sqrt{1+f'(R_0)} }{r_{\pm}}$. Clearly, the action growth rate of the static charged black holes in $f(R)$ gravity can also saturate Eq.~\eqref{conjectureofcai}. It means that the higher order derivative of the curvature term does not reduce the action growth rate of black holes within the WDW patch at the late time approximation limit. Though the contributions from the null surfaces and the joints were considered in Ref. \cite{Yangfr:2017}, the result is same with ours.

\subsection{Action growth rate of Kerr-Newman black holes in $f(R)$ gravity}

The KN solution for a constant curvature black hole in $f(R)$ gravity was found in Ref. \cite{Cembranos:2011sr}:
\begin{eqnarray}\label{metricoffr}
ds^2=&-&\left(\frac{\triangle _r}{\rho^2\Xi^2}-\frac{\triangle_\theta \sin^2\theta}
                                                {\rho^2\Xi^2}a^2\right)dt^2\nn\\
     &+&2\frac{a\triangle _r \sin^2\theta-a(r^2+a^2)\triangle_\theta \sin^2\theta}
            {\rho^2\Xi^2}dtd\phi\nn\\
     &+&\frac{\rho^2}{\triangle_r}dr^2
     +\frac{\rho^2}{\triangle_\theta}d\theta^2\nn\\
     &+&\frac{\sin^2\theta}{\rho^2\Xi^2}
       \left[(r^2+a^2)^2\triangle_\theta-a^2\triangle_r\sin^2\theta\right]d\phi^2,
\end{eqnarray}
where $a$ is the rotating parameter and
\begin{eqnarray}
\triangle _r&=&\left(a^2+r^2\right) \left(1-\frac{R_0}{12}r^2\right)
             +\frac{Gq^2}{1 +f'(R_0)}-2 Gmr,\nn\\
\triangle _{\theta }&=&1+\frac{ R_0}{12} a^2 \cos ^2\theta,\nn\\
\Xi &=&1+\frac{ R_0}{12}a^2,\nn\\
\rho ^2&=&r^2 +a^2 \cos ^2\theta.
\end{eqnarray}
Here we take the value of the Ricci scalar $R_0$ as a constant which is the same as Eq. \eqref{constant R}. We can solve the mass parameter $m$ and electric charge parameter $q$ from $\triangle_r(r_{\pm})=0$,
\begin{eqnarray}
m&=&\frac{\left(r_-+r_+\right) \left(a^2+l^2+r_-^2+r_+^2\right)}{2 G l^2},\nn\\
q^2&=&\frac{\left( 1 +f'(R_0)\right)}{G l^2} \Big(a^2 \left(r_- r_+-l^2\right)\nn\\
       &&+r_- r_+\left(l^2+r_-^2+r_- r_++r_+^2\right)\Big).
\end{eqnarray}
The electromagnetic field is described by
\begin{eqnarray}
A=&-&\frac{qr}{\rho^2}\left(\frac{dt}{\Xi}-a\sin^2\theta\frac{d\phi}{\Xi}\right),\nn\\
F=&-&\frac{q(r^2-a^2\cos^2\theta)}{\rho^4}
     \left(\frac{dt}{\Xi}-a\sin^2\theta\frac{d\phi}{\Xi}\right)\wedge dr\nn\\
  &-&\frac{2qra\cos\theta\sin\theta}{\rho^4}d\theta\wedge
      \left(a\frac{dt}{\Xi}-(r^2+a^2)\frac{d\phi}{\Xi}\right).
\end{eqnarray}
The action growth rate of the bulk part can be calculated as
\begin{eqnarray}
\frac{d\mathcal A_{bk}}{dt}
                           &=&\left(1+f'(R_0)\right)l^2\left(r_--r_+\right)
                                \frac{a^2+r_-^2+r_+^2+r_- r_+}
                                {2 G \left(l^2-a^2\right)^2}\nn\\
                           &&-\frac{l^4 q^2 \left(r_+-r_-\right) \left(a^2-r_- r_+\right)}
                          {2 \left(l^2-a^2\right)^2 \left(a^2+r_-^2\right)
                           \left(a^2+r_+^2\right)}.
\end{eqnarray}
For the boundary term, we take the same procedure as in GR. Using the condition $\triangle_r(r_\pm)=0$, we have
\begin{eqnarray}
\frac{d\mathcal A_{bd}}{dt}=&&\frac{\left(1 + f'(R_0)\right)l^2 \left(r_+-r_-\right)}
                              {2 G \left(l^2-a^2\right)^2}\nn\\
                             && \times\left(a^2+l^2+2\left(r_-^2+r_+ r_-+r_+^2\right)\right).
\end{eqnarray}
The angular velocity $\Omega$, electric potential $\mu$, physical mass $M$, charge $Q$, and angular momentum $J$ are \cite{Cembranos:2011sr}
\begin{eqnarray}
\Omega_{\pm}&=&\frac{a}{a^2+r_{\pm}^2},\nn\\
\mu_{\pm}&=&\frac{qr_{\pm}}{\Xi (a^2+r_{\pm}^2)},\nn\\
M&=&(1+f'(R_0))\frac{m}{\Xi},\nn\\
Q&=&\frac{q}{\Xi},\nn\\
J&=&(1+f'(R_0))\frac{am}{\Xi^2}.
\end{eqnarray}
With the help of these quantities, we finally obtain
\begin{eqnarray}
\frac{d\mathcal A}{dt}=\left(M- \Omega _+J- \mu _+Q\right)-\left(M- \Omega _-J- \mu _-Q\right).
\end{eqnarray}
So we confirmed the conjecture of Ref. \cite{Cai:2016xho}.

In the limit $a\rightarrow 0$, the KN black hole recovers to the charged black hole which can be seen from the metrics \eqref{KNmetric} and \eqref{metricoffr}.  The angular velocity and angular momentum will also approach to zero. Therefore, the action growth rate can recover to the charged black hole case. Furthermore, in the neutral limit $Q\rightarrow 0$, the inner horizon will disappear, i.e., $r_-\rightarrow 0$. In this case, the term $\mu_+Q$ approaches to zero and the term $\mu_- Q$ approaches to $2M$, so the action growth rate approaches to $2M$, which recovers to the neutral case. These analyses are valid not only for GR but also for massive gravity and $f(R)$ gravity.

What we have done shows that all the action growth rates of neutral, charged, and KN black holes in $f(R)$ gravity can saturate the conjecture \eqref{conjectureofcai}, because what we have investigated is the case of constant curvature. From the action \eqref{action of f(R)} we can see that, the higher derivative term, i.e., the $f(R)$ term only contributes to a cosmology constant for the constant curvature case. In other words, the higher derivative term has no contribution to the action growth rate. Though the boundary term is changed, the result still can saturate the bound \eqref{conjectureofcai}. As for the other higher derivative cases, the results may be different, and this is an interesting question to be investigated.

\section{Conclusion and discussion}\label{conclusion}
The action of a stationary AdS black hole within the WDW patch has been related to the quantum complexity of a holographic  state. Following the procedure in Ref. \cite{Cai:2016xho}, we calculated the action growth rate of the static and stationary AdS black holes in massive gravity and $f(R)$ gravity within the WDW patch at the late time approximation.

For the massive gravity, we found that the action growth rate of the black hole does not depend on the parameter $k$. That is to say, no matter what the topology of the black hole is, the action growth rate will saturate the bound \eqref{conjectureofcai}. The action growth rate equals to $2M$ for a neutral static AdS black hole, and it is the difference value of  $M-\mu Q$ at the inner and outer horizons for a charged static AdS black hole. Though the structure of the growth rate does not change, the effect of the massive graviton terms will slow down the growth rate for the charged static AdS black holes in massive gravity which has the same mass and charge with the black holes in GR. But the massive graviton terms do not have effect on the neutral static AdS black holes. These results are consistent with that of Ref.~\cite{Pan:2016ecg}.

For the $f(R)$ gravity, we found that the bound \eqref{conjectureofcai} of the action growth rate does not change for neutral, charged, and KN AdS black holes. Because the $f(R)$ term  for the case of constant curvature only contributes to the action a cosmology term, so the results are expected.  Our results checked the conjecture proposed in Ref.~\cite{Cai:2016xho} and showed more abundant features on the action growth rate as well as the quantum complexity. For other higher derivative gravities, this conjecture should be checked in detail.

Though the topology of the spacetime studied in this paper has no influence on the action growth rate, the result may be different for other nontrivial topologically black hole spacetime. It is well known that there are many interesting effects of the nontrivial topology in black hole spacetime. For example, black holes with topologically nontrivial $S^2$-cycle have the same conserved charge at infinity as the Breckenridge-Myers-Peet-Vafa (BMPV) black hole but greater entropy than the BMPV black hole~\cite{Horowitz:2017fyg,Kunduri:2014iga}. As the action of the WDW patch of the black hole is dual to the quantum complexity of the quantum state, to study the influence of the nontrivial topology on the action growth rate will be a further test on the CA duality. And this may lead to a deeper understanding on the black hole interior. However, the spacetime of the black hole with nontrivial topology found in Refs. \cite{Horowitz:2017fyg,Kunduri:2014iga} is asymptotically flat, we have no idea on the dual quantum state of the flat spacetime, and hence on the quantum complexity. The solutions found in Ref.~\cite{Horowitz:2017fyg} can be generalized to six dimensions, and the spacetime is asymptotically  $AdS_3\times S^3$, however, the CFT dual of the bubbling geometry is still unknown \cite{Skenderis:2008qn,Giusto:2015dfa}. Anyway, this would be an interesting direction.

Besides, black hole interior has also been a mystery, we hope that we can learn more about this from the view point of quantum complexity.

\section{Acknowledgments}

This work was supported by the National Natural Science Foundation of China (Grants Nos. 11522541, 11375075, and 11675064), and the Fundamental Research Funds for the Central Universities (Grants Nos. lzujbky-2016-k04, lzujbky-2016-115, and lzujbky-2017-it69).

\end{document}